\documentstyle{mn}

\def\etal{{\it et al.\thinspace}}
\def\ie{{\it i.e.\ }}
\def\vech{{\bf J}}
\def\refindent{\par \noindent \hang}
\def\paper#1#2#3#4#5{\refindent #1, #2, #3, #4, #5}
\def\and{, }
\def\AJ{AJ}
\def\ApJ{ApJ}
\def\MN{MNRAS}

\title[Precession in triaxial potentials]
{On the probability of major-axis precession in triaxial ellipsoidal
potentials}

\author[Thomas, Vine \& Pearce]
{P. A. Thomas, S. Vine$^*$ and F. R. Pearce\\
Astronomy Centre, MAPS, University of Sussex, Brighton, BN1 9QH\\
$^*$Current address: Institute of Astronomy, Madingley Road,
Cambridge, CB3 0HA}

\begin{document}

\maketitle
\begin{abstract}
Orbits in triaxial ellipsoidal potentials precess about either the major
or minor axis of the ellipsoid.  In standard perturbation theory it can
be shown that a circular orbit will precess about the minor axis if its
angular momentum vector lies in a region bounded by two great circles
which pass through the intermediate axis and which are inclined with
minimum separation $i_T$ from the minor axis, where  $i_T={\rm
arctan}\big((B^2-C^2)/(A^2-B^2)\big)^{1/2}$ and $A$, $B$ and $C$ are the axis
ratios, $A\geq B\geq C$. We test the accuracy of this formula by
performing orbit integrations to determine $i_S$, the simulated turnover
angle corresponding to $i_T$.

We reach two principal conclusions: (i) $i_S$ is usually greater than
$i_T$, by as much as 12 degrees even for moderate triaxialities,
$A/1.2<B<C/0.8$. This reduces the expected frequency of polar rings.
(ii) $i_S$ is not a single, well-defined number but can vary by a few
degrees depending upon the initial phase of the orbit.   This
means that there is a reasonable probability for capture of gas onto
orbits which precess about both axes.  Interactions can then lead to
substantial loss of angular momentum and subsequent infall to the
galactic centre.
\end{abstract}


\section{Introduction}

Polar ring galaxies are systems in which a ring of gas (and/or young
stars) is seen to orbit about the major axis of an early-type galaxy. In
many cases it has been established that the host galaxy is rotating at
right-angles to the ring and this is sometimes taken to be part of the
definition of a polar ring system.  There are just 6 confirmed polar
rings but 27 good candidates and many more possibles---see Whitmore
\etal (1990) for a review.

The host galaxy in polar ring systems often appears to be an S0 but this
leads to theoretical problems: in an oblate axisymmetric system all orbits will
precess about the minor axis at a rate which is a function of radius
(typically the period is proportional to radius).  This differential
precession will cause the polar ring to fragment. This can be overcome
if a small degree of triaxiality is assumed as orbits whose angular
momentum is sufficiently close to the major axis will then precess
around the major rather than the minor axis.  In general accreted discs
of gas will have an arbitrary orientation.  In this case differential
precession about a symmetry axis will lead to dissipation and collapse
into the plane perpendicular to that axis.

The regions of phase space which lead to precession about the major or
minor axes can be investigated using perturbation theory in Hamiltonian
mechanics.  It is assumed that the unperturbed potential is spherically
symmetric and that the orbits are circular.  The Hamilton is then
expanded in spherical harmonics and the time-averaged perturbing
potential calculated (see, for example, Steiman-Cameron \& Durisen,
1984).  This results in an expression of the form
\begin{equation}\label{eqpertpot}
 <\Phi_1>\propto C_{20}(3\sin^2i-2)+6C_{22}\cos2\Omega\sin^2i
\end{equation}
where $\Omega$ and $i$ are the node and inclination of the orbit, as
illustrated in Figure~\ref{figorbpar}, and $C_{20}$ and $C_{22}$ are
constants.  Orbits will precess along lines $<\Phi_1>=$constant as shown
in Figure~\ref{figamprec}.  If the angular momentum of the
orbit, $\vech$, lies within the shaded region then it will precess about
the major axis, otherwise it will precess about the minor axis.
The dividing lines between the two regions are great circles which pass
through the intermediate axis and which are inclined at an angle
\begin{equation}\label{eqit}
 i_T={\rm arcsin}\left(C_{20}+2C_{22}\over C_{20}-2C_{22}\right)^{1\over2}
\end{equation}
to the minor axis.  Hence
the probability that a randomly inclined disk will precess around
the major axis is
\[
 f=1-{2i_T\over\pi}.
\]
This formula is extensively used in the theory of polar rings.  It is
generally accepted that dissipation will cause a disk of gas to settle
into a plane perpendicular to the axis about which it precesses. Even a
small triaxiality can lead to a reasonable probability for polar ring
formation.  For example taking an ellipsoidal density distribution with
axis ratios $a=1.01$, $b=1$ and $c=0.8$ gives $i_T\approx 77^\circ$ and
$f\approx0.15$.

\begin{figure}
 \centering
 \vspace{7cm}
 \caption{The orbital elements.}
 \label{figorbpar}
\end{figure}

\begin{figure}
 \centering
 \vspace{7cm}
 \caption{Precession trajectories for the angular momentum vector of
 near-circular orbits in a triaxial potential.}
 \label{figamprec}
\end{figure}

In this paper we test the validity of the perturbation theory by
calculating orbits in an ellipsoidal triaxial potential.  We find that
the measured transition angle, $i_S$, which divides precession about
the major and minor axes, is usually larger than the theoretical
value, $i_T$.  Also the boundary dividing the two regions of
precession about the major and minor axes is not sharp.  The
precession axis depends upon the phase of the orbit and $i_S$ can vary
by a few degrees.  The integration method and results of our
simulations are presented in the Section~2.  The conclusions are
summarised and discussed in Section~3.

\section{Integration method and results}

\subsection{The integration method}

The numerical code used in this paper is that described by Pearce \&
Thomas (1991).  It is a simple predictor-corrector method integrates
orbits to high accuracy and we refer the reader to this paper for details.
%

We use a potential
\[
 \phi=\ln\left({x^2\over A^2}+{y^2\over B^2}+{z^2\over C^2}\right)
\]
where $A\geq B\geq C$ and without loss of generality we can take $B=1$.
The time-averaged first-order perturbation potential takes the form of
Equation~\ref{eqpertpot} with
\begin{eqnarray*}
 C_{20}&={1\over A^2}+{1\over B^2}-{2\over C^2}\\
 C_{22}&={1\over2}\left({1\over B^2}-{1\over A^2}\right)\\
\end{eqnarray*}
which gives, on substitution in Equation~\ref{eqit},
\[
 i_T={\rm arcsin}\left({1\over C^2}-{1\over B^2}
      \over{1\over C^2}-{1\over A^2}\right)^{1\over2}
    ={\rm arctan}\left(B^2-C^2\over A^2-B^2\right)^{1\over2}.
\]
Note that this expression is different from that for a system with
ellipsoidal density contours, $\rho=\rho((x/a)^2+(y/b)^2+(z/c)^2)$, for
which
\[
 i_T={\rm arcsin}\left(b^2-c^2\over a^2-c^2\right)^{1\over2}.
\]
These two formulae agree for near-spherical systems but can differ by 10
degrees or more for moderate triaxialities.  For example $a=1.2$, $b=1$,
$c=0.8$ gives $i_T=42.1\,$degrees whereas $A=1.2$, $B=1$, $C=0.8$ gives
$i_T=53.6\,$degrees.

Particles are given an initial tangential velocity of $\sqrt{2}$ which
would lead to a circular orbit in a spherical potential, $A=B=C$.  The
subsequent behaviour depends both upon the initial orientation of the
plane of the orbit and on the orbital phase of the particle.  These are
labelled by the co-ordinates shown in Figure~\ref{figorbpar}.  $i$ is
the inclination of the orbit to the $x$-$y$ plane (also the angle
between the angular momentum vector, $\vech$, and the $z$-axis),
$\Omega$ is the longitude of the ascending node, and $\chi$ is the phase
measured from the position of the ascending node.  Because the radial
velocity is initially zero $\chi$ corresponds also to the phase of the
peri- or apogee of the orbit.  Note, however, that this phase is not
conserved as it would be in a Keplerian potential.

There are too many free parameters to be able to investigate them all in
depth so we choose to restrict $\vech$ to lie initially in the $x$-$z$
plane (\ie $\Omega=\pi/2$).  If the transition lines separating the two
regions of precession about the major- and minor-axes are great circles
(as in linear perturbation theory) then it is trivial to generalise our
results to arbitrary $\vech$.

\subsection{Results}

The measured transition angle, $i_S$, is a function of the initial
phase, $\chi$, as illustrated in Figure~\ref{fig3} for one particular
choice of axis ratios, $A=1.1$, $B=1.0$ and $C=0.9$. The minimum value
of $i_S$ occurs at $\chi=0$ and the maximum at $\chi=\pi/2$ with an
approximately sinusiodal variation between the two.  Because of this we
need only present results for these two extremes, and the maximum and
minimum transition angles for a range of triaxialities are given in
Figure~\ref{figrange}.  The effect in each case is to lower the boundary
of the shaded region in Figure~\ref{figamprec} and to smear it out over
a few degrees.

\begin{figure}
 \centering
 \vspace{7cm}
 \caption{The measured turnover angle, $i_S$, as a function of initial
 phase for the case $A=1.1$, $B=1.0$ and $C=0.9$.  The theoretical value,
 $i_T$, is shown as a dotted line.}
 \label{fig3}
\end{figure}

\begin{figure*}
 \vbox to200mm{}
 \caption{}
 \label{figrange}
\end{figure*}

As an aside, we never see orbits which switch from precessing about the
major to the minor axis, or vice versa.  In general the phase of the
apogee of the orbit is not conserved.  However for those orbits which
lie close to the transition angle, $i_S$, the phase returns almost
exactly to its original value after one whole precession time.  We do
not know why this is the case although it is presumably a reflection of
some underlying conservation law.

\section{Conclusions}

In this paper we calculate the simulated transition angle for precession
about the major or minor axis of a triaxial potential and reach the
following two conclusions: (i) $i_S$ is almost always greater than $i_T$
and (ii) $i_S$ is spread out over a few degrees depending upon the
initial phase of the orbit.

The first of these results means that theoretical estimates, based on $i_T$,
of the expected fraction of accreted gas disk which will settle down to
give polar rings will be too high.   However the error is not likely to be
greater than about 10 percent, much lower than the uncertainty in the
observed frequency of polar ring systems.

More interesting is the second result which may provide a mechanism
for overcoming the angular momentum barrier which prevents accretion
into the core of a galaxy.  If material is accreted at an inclination
between the measured maximum and minimum values of $i_S$, and if the
accreted material is spread out over a range of phases, then
precession will occur around both the minor and the major axes.
Interactions between the two components can then lead to a large
reduction in angular momentum.  Indeed, once each component has
precessed through 180 degrees then the planes of their orbits again
coincide but their angular momenta are oppositely aligned.  Collisions
between gas clouds of similar mass may then reduce their velocity
relative to the galactic centre to almost zero and they will be
accreted into the core.  This situation may not be as unlikely as it
at first appears because the appropriate range of $i_S$ may span up to
10 degrees even for moderate triaxialities.  Also accretion of a
gas-rich dwarf spiral or dwarf irregular galaxy is likely to populate
a fair region of phase space and a large amount of dissipation is
required for this to settle into a disk of gas clouds on circular
orbits, as is often assumed for simplicity.

\section*{Acknowledgements}

We acknowledge the facilities of the STARLINK minor node at Sussex.

\section*{References}

\paper{Pearce, F. R.\and Thomas, P. A.}{1991}{\MN}{248}{688}

\paper{Steiman-Cameron, T. Y.\and Durisen, R. H.}{1984}{\ApJ}{276}{101}

\paper{Whitmore, B. C., Lucas, R. A., McElroy, D. B., Steiman-Cameron, T. Y.,
Sackett, P. D.\and Olling, R. P.}{1990}{\AJ}{100}{1489}

\end{document}